\documentstyle[11pt,twoside,psfig,jltp]{article}

\title{Scaling of the Equilibrium Magnetization in the Mixed State of Type-II Superconductors}

\author{I. L. Landau \address{Laboratorium f\"ur 
Festk\"orperphysik, ETH H\"onggerberg, CH-8093 Z\"urich, 
Switzerland} and H. R. Ott }

\runninghead{I. L. Landau and H. R. Ott}{Scaling of the Equilibrium Magnetization }
       
\begin{document}

\begin{abstract}
We discuss the analysis of mixed-state magnetization data of type-II superconductors using a recently developed scaling procedure. It is based on the fact that, if the Ginzburg-Landau parameter $\kappa$ does not depend on temperature, the magnetic susceptibility $\chi (H,T)$ is a universal function of $H/H_{c2}(T)$, leading to a simple relation between magnetizations at different temperatures. Although this scaling procedure does not provide absolute values of the upper critical field $H_{c2}(T)$, its temperature variation can be established rather accurately. This provides an opportunity to validate theoretical models that are usually  employed for the evaluation of $H_{c2}(T)$ from equilibrium magnetization data. In the second part of the paper we apply this scaling procedure for a discussion of the notorious first order phase transition in the mixed state of high-$T_c$ superconductors. Our analysis, based on experimental magnetization data available in the literature, shows that the shift of the magnetization accross the transition may adopt either sign, depending on the particular chosen sample. We argue that this observation is inconsistent with the interpretation that this transition always represents the melting transition of the vortex lattice.

\end{abstract}

\maketitle


\section{INTRODUCTION}

Measurements of the equilibrium magnetization  in the mixed state of a type-II superconductor provide valuable information about different parameters characterizing the superconducting state of the investigated material. This is particularly true for high-$T_c$ (HTSC) superconductors, because of the extremely wide range of magnetic fields in which their magnetization is reversible.\cite{irr} The physically meaningful information is not straightforwardly accessible, however, and theoretical models have to be invoked in analyses of experimental data. In this short review we discuss a recently developed scaling procedure and its application to the analysis of equilibrium magnetization measurements in the mixed state of type-II superconductors.\cite{1,6,cm} The comparison of our results with those obtained earlier by using alternative methods of analyzing the data shows that some of the assumptions that are made  in traditional theoretical approaches are not justified. For instance, we show that the impact of thermal fluctuations on the equilibrium magnetization is much weaker than previously claimed. We also apply the scaling approach in analyzing the well known first order phase transition in the mixed state of HTSC's which is usually attributed to the melting of the vortex lattice.\cite{tr1,tr2,tr3,tr4,tr5,tr6,tr7,tr9} 

In order to illustrate the problems that arise in the interpretation of the experimental data, we first consider several commonly used approaches for establishing the upper critical field and its temperature dependence $H_{c2}(T)$ from magnetization $M(T)$ data. 

\subsection{Intuitive approach}

 As an example we pick a study of an optimally doped YBa$_2$Cu$_3$O$_{7-x}$ (Y-123) sample that was published in Ref. \citen{welp1}. First, we consider the magnetization measurements in very low magnetic fields (Fig. 1(a)). The shape of this low-field $M(T)$ curve is exactly as expected for a type-II superconductor with non-zero pinning. In this case, the width of the observed transition is not connected with the sample non-uniformity but rather reflects a decrease of the critical current density by approaching $T_c$. The zero-field critical temperature $T_c$ can be reliably evaluated as that value of temperature, at which the diamagnetic moment of the sample vanishes. 

\begin{figure}[h]
\centerline{\psfig{file=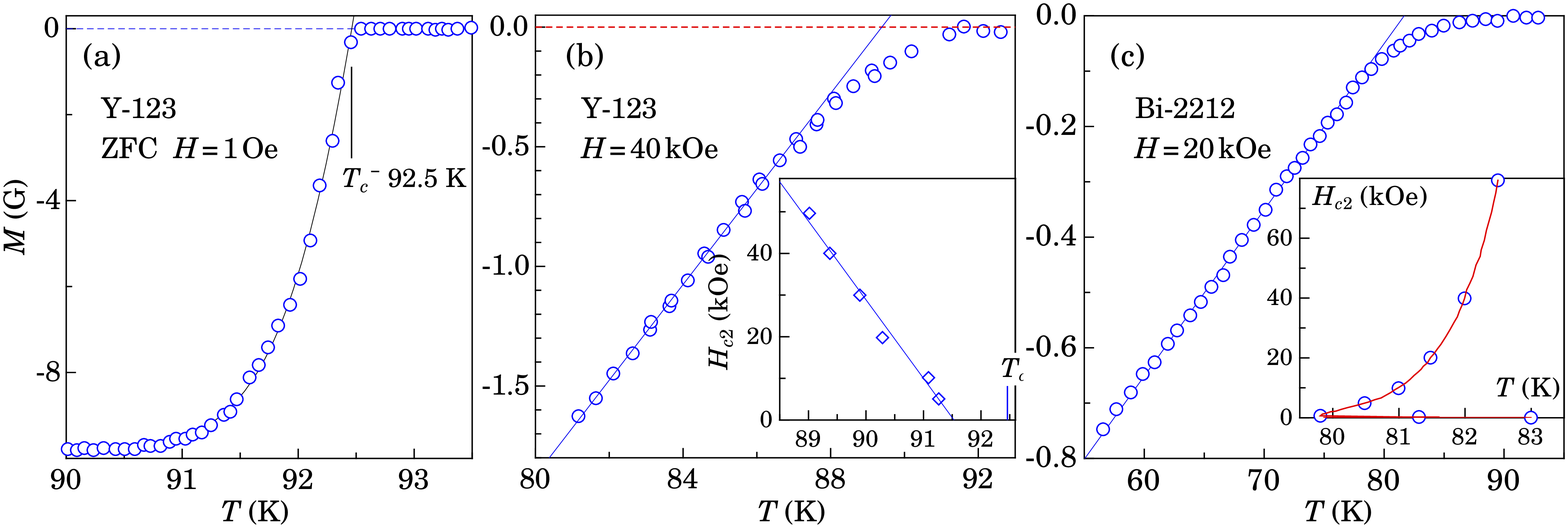,height= 1.7 in}}
\caption{Magnetization curves for Y-123 and Bi-2212 samples studied in Ref. \protect\citen{welp1} and  \protect\citen{krit}, respectively.  (a) A zero-field-cooled  (ZFC) $M(T)$ curve. The solid line is a guide to the eye.  (b) and (c)  Reversible magnetization data for two different samples. The solid lines represent linear approximations to the low temperature part of the curves. The insets show the resulting behavior of $H_{c2}(T)$. Note the difference in $T_c$ mentioned in the text in the inset of panel (b). }
\end{figure}
In higher magnetic fields (Fig. 1(b)), however, the situation is more complicated and the corresponding value of the critical temperature in a given external field, $T_{c2}(H)$, cannot be evaluated in the same simple manner as the zero-field $T_c$. In order to evaluate $H_{c2}(T)$, some {\it a priory} assumptions have to be made. In Ref. \citen{welp1}, it was postulated that some rounding of the $M(T)$ curve above $T \approx 88$ K is due to thermal fluctuations. According to arguments presented in Ref. \citen{welp1},\cite{fn0} the corresponding value of $H_{c2}(T)$ may be evaluated as the intersection point of the extrapolation of the linear part of the $M(T)$ curve with the abscissa axis (see Fig. 1(b)). As expected, $H_{c2}$ evaluated in such a way varies linearly with temperature (see the inset of Fig. 1(b)). At the same time, the value of $T$, at which $H_{c2}(T)$ vanishes, is considerably lower than $T_c$, evaluated as shown in Fig. 1(a). Although this situation casts some doubts on the validity of the assumptions that were made, the same approach was frequently used for evaluations of $H_{c2}(T)$ of Y-based cuprates.\cite{welp2,welp3,welp4,welp5}

The discrepancies are even worse if we apply the same approach to Bi-based HTSC's. The corresponding plots for a Bi$_2$Sr$_2$CaCu$_2$O$_8$ (Bi-2212) sample, studied in Ref. \citen{krit}, are shown in Fig. 1(c). Although the shape of the $M(T)$ curve for this sample is practically identical to that shown in Fig. 1(b), the resulting temperature dependence of $H_{c2}(T)$, shown in the inset of Fig. 1(c), demonstrates rather clearly the failure of the approach. In Ref. \citen{krit}, this failure was claimed to be due to an especially important role of thermal fluctuations in layered cuprates. Although this explanation is plausible, it should be considered as an assumption rather than a firm conclusion and, of course, this assumption does not help to establish $H_{c2}(T)$. 

\subsection{Hao-Clem model and thermal fluctuations of vortices}

Another widely used approach to extract $H_{c2}(T)$ from experimental magnetization data is by employing the model of Hao and Clem.\cite{haocl} This model is an analytical approximation to the Ginzburg-Landau theory of the Abrikosov vortex lattice. It allows to find the values of the Ginzburg-Landau parameter $\kappa$ and the thermodynamic critical magnetic field $H_c$ for a given temperature by fitting theoretical $M(H)$ curves to isothermal magnetization measurements. An important advantage of this approach is that equilibrium magnetization data collected in magnetic fields well below $H_{c2}$ may be used and therefore the accessible $H_{c2}(T)$ curve is not limited to temperatures very close to $T_c$.

\begin{figure}[h]
\centerline{\psfig{file=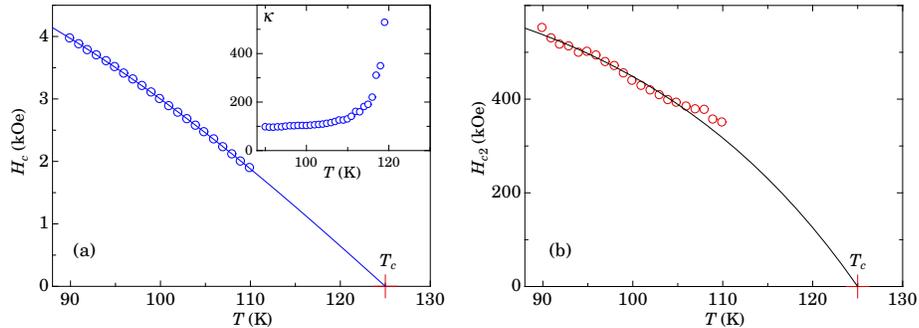,height= 1.75 in}}
\caption{(a) $H_c(T)$ and $\kappa(T)$ (inset) for a Hg-based sample studied in Ref. \protect\citen{haocl1}. (b) The $H_{c2}(T)$ curve calculated as $H_{c2} = \sqrt{2} \kappa H_c$. The solid  lines are guides  to the eye.}
\end{figure}
As a first example, we  consider the results for a Hg-based sample published in Ref. \citen{haocl1} (Fig. 2). While the shape of the $H_{c}(T)$ curve appears as reasonable, the temperature dependence of $\kappa$ reveals a strong and unphysical increase of $\kappa$ with increasing temperature. This type of $\kappa(T)$ behavior resulted in practically all published analyses, in which the Hao-Clem model was applied to experimental magnetization data.\cite{haocl2,haocl3,haocl4,haocl5,haocl6,haocl7,haocl8,haocl9,haocl9a,haocl10,haocl11,haocl12,haocl13} The standard explanation is that because of thermal fluctuations of vortices, which are not accounted for by the model, this approach is inapplicable at higher temperatures. Considering the data presented in Fig. 2(b), we may conclude that for this particular sample the Hao-Clem analysis is valid only at temperatures $T \le 105$ K $= 0.84T_c$. In some other cases, however, the same type of analysis leads to manifestly wrong results even at lower temperatures. Two such examples are shown in Fig. 3. Contrary to common expectations, $H_{c2}(T)$ does not decrease with increasing temperature. 
\begin{figure}[h]
\centerline{\psfig{file=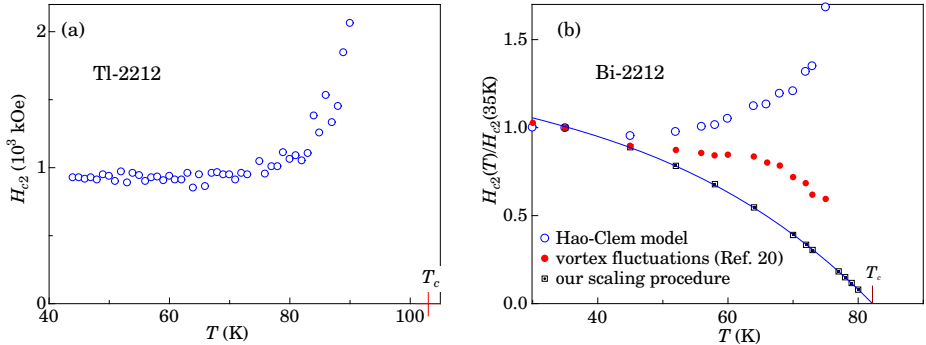,height= 1.85 in}}
\caption{The $H_{c2}(T)$ curves resulting from Hao-Clem type analyses for Tl$_2$Ba$_2$CaCu$_2$O$_{8+x}$ (Tl-2212)  and Bi-2212 samples, studied in Refs. \protect\citen{haocl13} and \protect\citen{haocl4}, respectively. For the latter sample, calculations based on the theory developed in Ref. \protect\citen{kogan} and our method, which is described below, are also shown. The solid line in frame (b) is a guide to the eye. }
\end{figure}

In order to explain these failures of the model, two different theories were developed.\cite{kogan2,kogan} In Ref. \citen{kogan2},  it was assumed that at low temperatures $T \ll T_c$, the discrepancies are due to effects of non-locality. The steep increase of $H_{c2}$ at higher temperatures was ascribed to extremely strong fluctuation effects in these HTSC compounds.\cite{kogan} Indeed, if magnetization data are treated by taking into account fluctuation effects,  the resulting $H_{c2}(T)$ curve adopt a more reasonable shape (see Fig. 3(b)).

Thermal fluctuations are undoubtedly important in HTSC's at temperatures close to $T_c$. For instance, they account for the experimental observation that in layered  HTSC compounds, the derivative $dM/dH$ changes sign at one particular temperature in all magnetic fields.\cite{blk} The rather ad hoc assumption of their importance, simply to correct for the inadequacies of the above mentioned model, is less convincing, and it may well be that the failure of the model is due to completely different reasons. Several such possibilities are listed below.

The model's three main assumptions include

(i) Conventional $s$-wave pairing. It is, however, commonly accepted that $d$-wave pairing is responsible for superconductivity in cuprates. The free energy includes a term describing the spatial distribution of the superconducting order parameter and therefore the equilibrium configuration of the mixed state, which is determined by a minimum of the free energy, is sensitive to the pairing type.

(ii) The sample geometry. The model assumes an infinite extension of the sample in the direction of the applied magnetic field. Because real HTSC samples are often rather thin, the free energy corrections due to a non-uniform magnetic field distribution outside the sample should also be taken into account. 

(iii) The normal state background. It is assumed that the normal-state magnetic susceptibility $\chi^{(n)}$ is negligible. This is not the case in HTSC's. Because $\chi^{(n)}$ is temperature dependent, it is difficult to identify the corresponding contribution from experimentally measured magnetizations. 

Summarizing, we note that none of the main assumptions of the Hao-Clem model is actually satisfied in real experiments. Any disagreement between a theoretical $M(H)$ curve and corresponding experimental data is thus not surprising at all. Below we consider a simple scaling procedure which allows for the evaluation of the temperature dependence of a normalized upper critical field from magnetization measurements. This procedure is based on the single assumption that the Ginzburg-Landau parameter $\kappa$ is temperature independent. The resulting temperature dependencies of $H_{c2}$ turn out to be quite different from those quoted above.

\section{THE SCALING PROCEDURE}

If the Ginzburg-Landau parameter $\kappa$ is temperature independent,\cite{fn1} the magnetic susceptibility $\chi$ in the mixed state of a type-II superconductor may be written as 
\begin{equation}
\chi(H,T) = \chi(H/H_{c2}),
\end{equation}
According to Eq. (1), the temperature dependence of $\chi$ is solely determined by the temperature variation of the upper critical field. Eq. (1) is sufficient to establish a relation between the magnetizations $M$ at two different temperatures $T$ and $T_0$, which may be written as 
\begin{equation}
M(H/h_{c2},T_0) = M(H,T)/h_{c2}
\end{equation}
where $h_{c2}(T)  = H_{c2}(T)/H_{c2}(T_0)$ is the ratio of the upper critical fields at $T$ and $T_0$. This equation is valid if the diamagnetic response of the mixed state is the only significant contribution to the sample magnetization. It is well known, however, that many superconducting materials, including HTSC's, exhibit sizable paramagnetic susceptibilities in the normal state. In order to account for this additional contribution to the sample magnetization, the following modification of Eq. (2) needs to be made\cite{1} 
\begin{equation}
M(H/h_{c2},T_0) = M(H,T)/h_{c2} + c_0(T)H,
\end{equation}
with
\begin{equation}
c_0(T)=\chi_{eff}^{(n)}(T)-\chi_{eff}^{(n)}(T_0),
\end{equation}
where $\chi_{eff}^{(n)}(T)$ is the effective magnetic susceptibility of the superconductor in the normal state. In practice and depending on the measurement technique, $\chi_{eff}^{(n)}$ may also include a non-negligible contribution arising from the sample holder. In the following we use $M_{eff}(H)=M(H,T_0)$ to denote the field dependence of the magnetization at $T = T_{0}$, calculated from measurements at $T \ne T_{0}$ using Eq. (3). The adjustable parameters $h_{c2}(T)$ and $c_{0}(T)$ are obtained by sticking to the condition that the $M_{eff}$ curves, calculated from measured $M(H)$ data in the reversible regime at different temperatures, collapse onto a single $M_{eff}(H)$ curve, which represents the equilibrium magnetization at $T = T_{0}$ (see Ref. \citen{1} for details). 

There are two adjustable parameters in our scaling procedure. The upper critical field $H_{c2}$ represents the natural normalization parameter for all magnetic characteristics of the mixed state and, as stated above, $c_0(T)$ is essential to account for any temperature dependence of the magnetic susceptibility in the normal state. 

The condition expressed by Eq. (1) is also satisfied in the above mentioned model of Hao and Clem.\cite{haocl}  An important advantage of our scaling approach, however, is that no particular field dependence of the magnetization has to be assumed and therefore, this procedure is free from those limitations that we outlined at the end of the previous section. The scaling procedure can be used for any type-II superconductor, independent of the pairing type, the absolute value of $\kappa$, the anisotropy of superconducting parameters or the sample geometry. In addition, it is equally valid for different mixed state configurations such as a vortex lattice, a vortex liquid, a vortex glass, or even a configuration of superconducting filaments in a normal matrix, considered in Ref.  \citen{fil}. It is only important that the magnetization curves that are used in the analysis are reversible, i.e., represent the equilibrium magnetization.  

\section{SCALING ANALYSIS OF MAGNETIZATION DATA. TEMPERATURE DEPENDENCE OF $H_{c2}$}

As a first example, we consider the magnetization data for a Tl-2212 sample presented in Ref. \citen{haocl13}. The dependence of $M_{eff}$ on $H/h_{c2}(T)$ that results from our scaling procedure is shown in Fig. 4. Because of the high quality of the original experimental data and the extended covered range of magnetic fields, the $M_{eff}(H/h_{c2})$ data points, calculated from the measurements at different temperatures, combine to a single curve with virtually no scatter.
\begin{figure}[h]
\centerline{\psfig{file=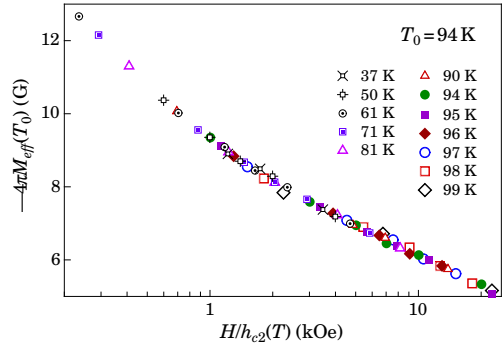,height= 1.85 in}}
\caption{The scaled magnetization $M_{eff}$ calculated for the Tl-2212 sample studied in Ref. \protect\citen{haocl13}.  }
\end{figure}

The temperature dependence of the scaling parameter $c_0$, which is shown in Fig. 5(a), demonstrates that the paramagnetic contribution to the sample magnetization obeys a Curie-type law in a rather extended temperature range with some deviations at temperatures below 55 K, as well as at temperatures very close to $T_c$.

\begin{figure}[h]
\centerline{\psfig{file=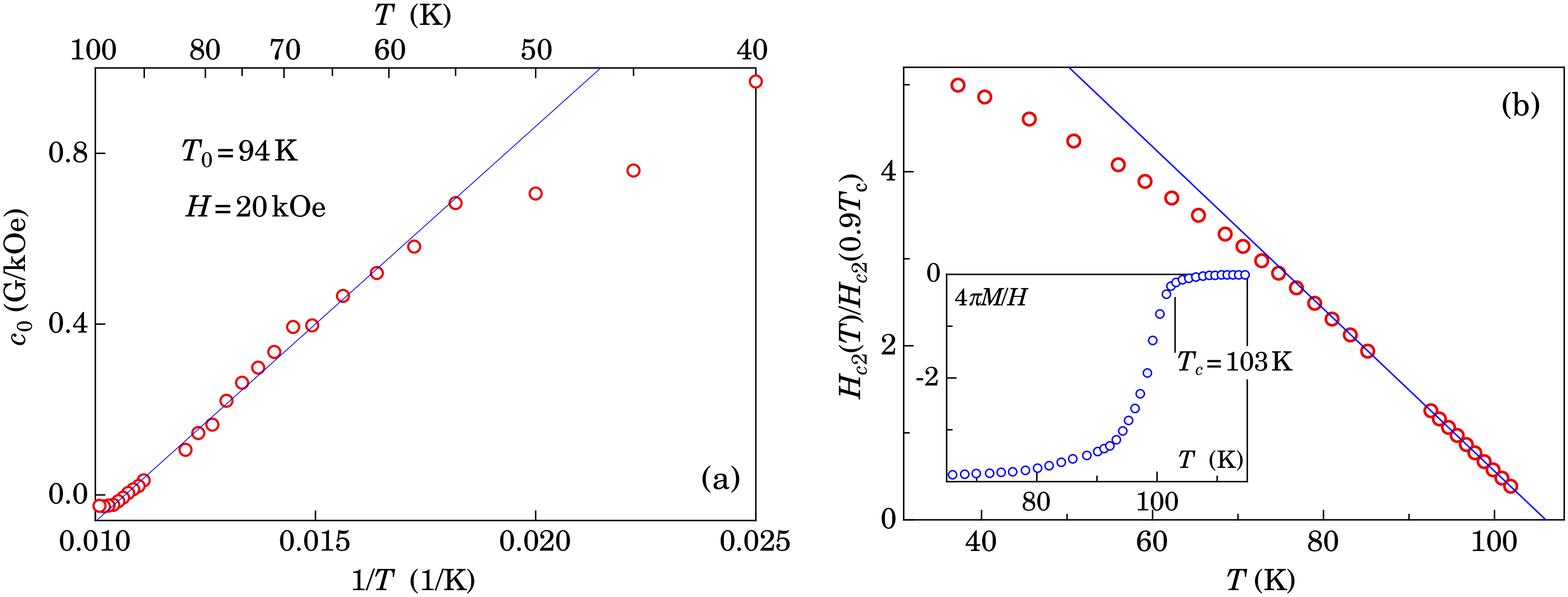,height= 1.85 in}}
\caption{(a) The scaling parameter $c_0$ as a function of $1/T$. The solid line represents the Curie-Weiss type behavior of $\chi ^{(n)}(T)$. (b) The temperature dependence of the normalized upper critical field. The solid line is the best linear fit to the data points for $T \ge 80$ K. The inset shows the $M(T)$ data measured at $H = 10$ Oe.\protect\cite{haocl13} The short vertical line indicates the position of $T_c$ as evaluated by the linear extrapolation of $h_{c2}(T)$ to $h_{c2} = 0$. }
\end{figure}
The normalized upper critical field $h_{c2}(T)$ is shown in Fig. 5(b). As may be seen, $h_{c2}(T)$ for this sample varies linearly with $T$ above $T \approx 80$ K. This linearity allows for a quite accurate evaluation of the critical temperature $T_{c}$ by extrapolating the $h_{c2}(T)$ curve to $h_{c2} = 0$. The inset of Fig. 5(b) demonstrates that the value of $T_{c} = 103.0$ K is well in agreement with the temperature dependence of the low-field magnetization of the same sample. We note that the resulting temperature dependence of $H_{c2}$ is significantly different from the results of previous analy§ses of the same data, shown in Fig. 3(a). 

The results of an analogous analysis of data for a Bi-2212 sample, reported in Ref. \citen{haocl4}, are shown in Fig. 3(b). As may be seen, our $h_{c2}(T)$ curve is again rather different from those obtained by other approaches. The temperature dependence of $H_{c2}$, as it follows from our analysis, is well in agreement with expectations: it is linear close to $T_c$, with a trend to saturation at lower temperatures.

Because the influence of thermal fluctuations on the magnetization of HTSC's in the mixed state is often overestimated in the literature, we discuss this point in more detail. As may clearly be seen in Fig. 4, if the experimental magnetizations are corrected for the temperature dependence of $\chi_{eff}^{(n)}$, the 
$M_{eff}(H)$ data points, calculated from data that were obtained at different temperatures, merge onto exactly the same curve. The obvious conclusion is that fluctuation effects do not contribute to the sample magnetization in this rather wide temperatures range. However, a close inspection of Fig. 5(a) reveals that the $c_0(T)$ data points for the three highest temperatures ($T \ge 97$ K) deviate from the straight line which represents the Curie law. It is plausible that these deviations are indeed due to thermal fluctuations. As was pointed out in Ref. \citen{2}, at temperatures close to $T_c$, the term $(c_0(T)H)$ may also account for fluctuation effects. However, because the fluctuation-induced contribution to the sample magnetization is not linear in $H$, it may only approximately be accounted for.  At still higher temperatures $T \ge 100$ K $\approx 0.97T_c$, the magnetization data presented in Ref. \citen{haocl13} cannot satisfactorily be scaled using Eq. (3). We consider these observations as evidence for the increasing role of fluctuations with increasing temperature, but only close to $T_c$. For Y-123 compounds, the impact of fluctuations effects is even weaker and in some cases our scaling procedure could successfully be employed up to temperatures as high as $0.99T_c$.\cite{1} 
\begin{figure}[h]
\centerline{\psfig{file=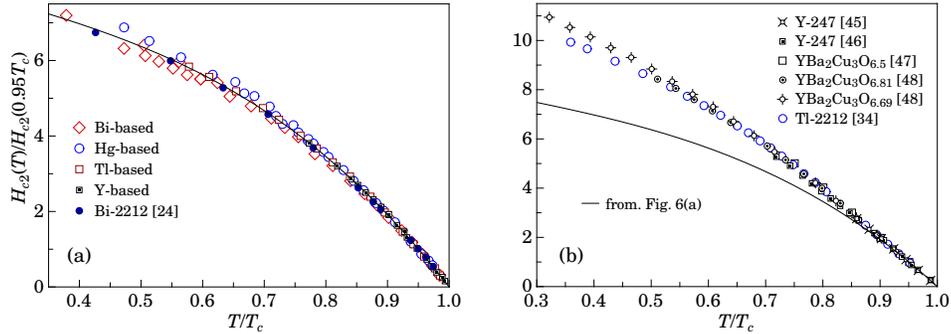,height= 1.8 in}}
\caption{$H_{c2}(T)/H_{c2}(0.95T_c)$ versus $T/T_c$ for two groups of HTSC's (see text for details).  In both frames, the solid line represents the same data set.}
\end{figure}

In order to compare the results obtained for different samples of HTSC's, we plot $H_{c2}(T)/H_{c2}(0.95T_c)$ versus $T/T_c$ in Fig. 6. It turns out that all the numerous HTSC's may be divided into two groups with two distinctly different normalized $H_{c2}(T)$ curves. The corresponding results are shown in Fig. 6. As may be seen in Fig. 6, the Bi-2212 and Tl--2212 samples that we discussed above belong to different groups. Note that the first group (Fig. 6(a)) includes a great variety of different HTSC compounds, while the second one (Fig. 6(b)) is apparently limited to a few Y- and Tl-based materials (see Refs. \citen{2} and \citen{6} for details). The universality of the $H_{c2}(T)$ curves for different HTSC's is undoubtedly the most surprising result of our analysis so far. It is difficult to imagine that this universality is simply accidental. The spectacular coincidence of the $h_{c2}(T/T_{c})$ data for a great variety of different samples is rather an unambiguous evidence that our approach captures the essential features of the magnetization process of HTSC's. It does not necessarily mean, however, that the Ginzburg-Landau parameter $\kappa$ is indeed temperature independent. The universality of $h_{c2}(T/T_{c})$ is preserved if the temperature dependence of $\kappa$ is the same for different HTSC compounds. Because the temperature dependence of $\kappa$ is weak,\cite{bcs} it does not change the  $h_{c2}(T)$ curves significantly.\cite{4}

\section{SCALING ANALYSIS OF PHASE TRANSITIONS IN HTSC'S}

In this section, we consider the scaled magnetization curves $M_{eff}(H)$ in the vicinity of the well-known first order phase transition in the mixed  state of HTSC's. The analysis of such curves provides some new insight in the very complex physics of the phase transition. In existing literature this phase transition is almost always attributed to vortex-lattice melting. However, some of the results of our following analysis are in contradiction with the vortex-lattice-melting hypothesis. Hence, in order to distinguish the two phases above and below the transition, respectively, we introduce the notion of "high-field" and "low-field phase", instead of the commonly used notations of vortex liquid and vortex solid. We divide this section into two parts. In the first subsection, we present and discuss the analysis of the isothermal $M(H)$ curves, while in the second  we demonstrate that the scaling procedure may also be used for the scaling of $M(T)$ curves measured in fixed magnetic fields.

\subsection{Analysis of $M(H)$ curves}

First, we consider the experimental data for two similar La$_{1.908}$Sr$_{0.092}$CuO$_4$ (La-214) single crystals, experimentally investigated in Refs. \citen{sasa1} and \citen{sasa2}. The scaling results for these samples are shown in Fig. 7(a). In these two experiments, the magnetization is reversible down to magnetic fields well below the phase transition. The collapse of the data points measured at different temperatures is achieved with the same values of $h_{c2}(T)$ and $c_0(T)$ both below and above the transition, in complete agreement with the expectation outlined above. The results displayed in Fig. 7(a) may be interpreted as evidence for the existence of two different modifications of the mixed state with two different equilibrium $M_{eff}(H,T_0)$ curves above and below the transition. In the following we use $M_{eq}^{(l)}$ and $M_{eq}^{(h)}$ to distinguish between the equilibrium $M_{eff}(H)$ curves for the low- and the high-field modifications of the mixed state, respectively. As may be seen in Fig. 7(a), the low-field modification corresponds to slightly higher values of the diamagnetic moment. The difference $\Delta M = {\left| {M_{eq}^{(l)}} \right|} - {\left| {M_{eq}^{(h)}} \right|}$ is positive. 

\begin{figure}[h]
\centerline{\psfig{file=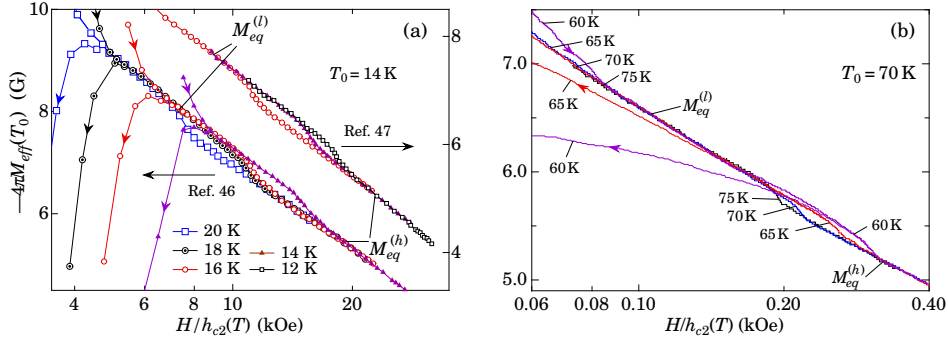,height= 1.8 in}}
\caption{(a) $M_{eff}(H/h_{c2})$ in the vicinity of the transition for two La-214 samples studied in Refs. \protect\citen{sasa1} and \protect\citen{sasa2}. For the sample of Ref. \protect\citen{sasa2}, only reversible magnetization data are shown. The analogous diagram for a Bi-2212 sample studied in Ref.  \protect\citen{kimurar} is displayed in (b).}
\end{figure}
The results for the Bi-2212 sample investigated in Ref. \citen{kimurar} are shown in Fig. 7(b). The $M(H)$ curves measured at $T=75$ K and $T=70$ K are reversible in the entire covered field range.\cite{kimurar} As for the La-214 samples, the $M_{eff}(H,T_0)$ curves collapse, both above and below the transition, clearly indicating  $M_{eq}^{(l)}(H)$ and $M_{eq}^{(h)}(H)$. The unexpected difference to the results shown in Fig. 7(a) is that the branches of the magnetization curves measured in increasing fields follow the equilibrium magnetization $M_{eq}^{(l)}$ curve already at magnetic fields well below the irreversibility line. This behavior seems to be common for Bi-based HTSC's and another example is shown in Fig. 8(a).\cite{5} 

\begin{figure}[h]
\centerline{\psfig{file=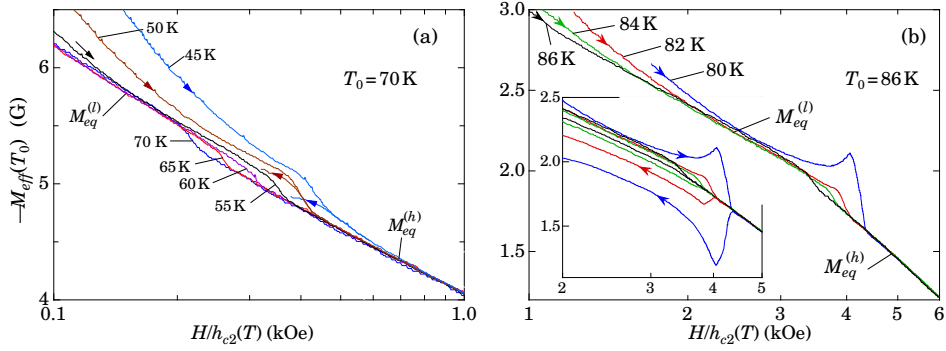,height= 1.85 in}}
\caption{The scaled magnetization in the transition region for (a) a Bi-2212 sample investigated in Ref.   \protect\citen{kimura}, and (b) an Y-123 sample investigated in Ref. \protect\citen{koba}. The inset in frame (b) shows both branches of $M(H)$ for the same temperatures as in the mainframe.}
\end{figure}
The behavior of the magnetization of an optimally doped Y-123 sample in the vicinity of the phase transition is similar to that for Bi-based samples. In spite of a pronounced peak effect at the lowest temperature, the $M_{eq}^{(l)}$ and $M_{eq}^{(h)}$ curves are readily identified. 

The results presented in Figs. 7(b) and 8 demonstrate that in Bi-2212 and Y-123 samples, the effect of pinning is strongly dependent on the direction of the flux motion. The pinning effects are obviously much weaker for the magnetic flux entering the sample. One of the possible explanations of this pinning force asymmetry was considered in Ref. \citen{5}. Note that this asymmetry is practically absent in the data for the La-214 samples, shown in Fig. 7(a).

The $M_{eff}(H)$ curves in the transition region for another La-214 sample, which are displayed in Fig. 9(a), are quite different from those shown in Figs. 7 and 8. The easily distinguishable $M_{eq}^{(l)}(H)$ and $M_{eq}^{(h)}(H)$ curves indicate that the difference $\Delta M = {\left| {M_{eq}^{(l)}} \right|} - {\left| {M_{eq}^{(h)}} \right|}$ is negative. The reason for the sign difference of $\Delta M$ between the data shown in Fig. 7(a) and those in Fig. 9(a) is difficult to establish. According to the original work,\cite{sasa1,sasa2,sasa3} the three La-214 samples are practically identical. 

Finally, in Fig. 9(b), we show the result of the scaling procedure for an overdoped Bi-2212 sample studied in Ref. \citen{sasa3}. Although only $M(H)$ data in magnetic fields above the transition were used for the evaluation of $h_{c2}(T)$ and $c_0(T)$, the magnetization curves measured in increasing magnetic fields collapse onto a single $M_{eff}(H)$ curve also below the irreversibility line and thus, as 
argued above, represent the equilibrium magnetization for $T = T_0$ in this field range.\cite{5} The equilibrium $M(H)$ curves for the low and high field modifications of the mixed state are such that the difference $\Delta M$ in the transition region is again negative. The width of the transition from one modification of the mixed state to the other, which is quite large at higher temperatures, is considerably reduced with decreasing temperature. 

\begin{figure}[!h]
\centerline{\psfig{file=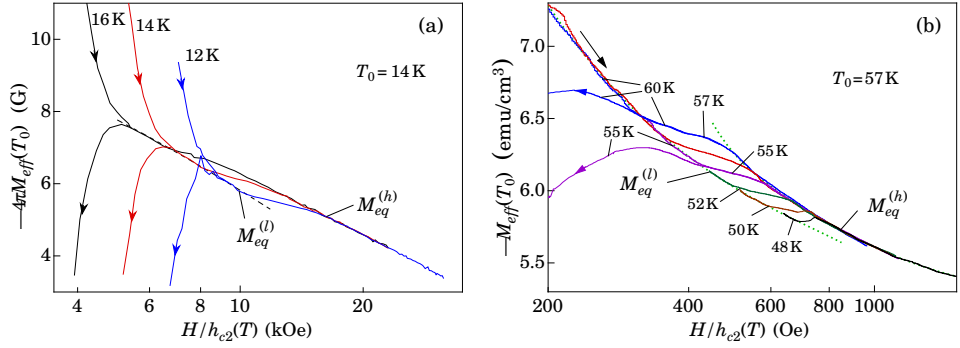,height= 1.85 in}}
\caption{(a) $M_{eff}(H)$ for a La-214 sample studied in Ref. \protect\citen{sasa3}. The dashed line indicates $M_{eq}^{(l)}(H)$. (b) $M_{eff}(H)$ for an overdoped Bi-2212 sample.\protect\cite{hana} The dotted lines indicate $M_{eq}^{(l)}(H)$ and $M_{eq}^{(h)}(H)$, respectively. Because of strong fluctuations in part of the magnetization data, shown in Fig. 2(a) of Ref. \protect\citen{hana}, the corresponding branches of the $M_{eff}(H/h_{c2})$ curves are not displayed in this plot.}
\end{figure}
The most unexpected result of our analysis is that the magnetization difference $\Delta M$ across the transition may adopt either sign. This result is difficult to reconcile with the vortex-lattice-melting 
hypothesis. In case of vortex lattice melting, the external magnetic field acts as pressure does in traditional solid-liquid melting transitions. Thermodynamics requires that the phase corresponding to the 
higher pressure must have a higher density, independent of whether this high-pressure phase is a liquid or a solid. In relation with the mixed state of type-II superconductors, the vortex liquid necessarily has to adopt a higher vortex density, i.e., the difference $\Delta M$ must always be positive. Since negative values of $\Delta M$ are identified for materials belonging to two different families of HTSC's, this can hardly be refuted as an accidental result. 

In the bulk of the existing literature (see, for instance, Refs. \citen{tr1,tr2,tr3,tr4,tr5,tr6,tr7,tr9}), the first order transition in the mixed state of HTSC's is viewed as a melting transition in the system of vortices. In this scenario, the mixed state above the transition represents the vortex liquid, while the vortex solid below the transition is described as a lattice or Bragg glass of vortices, depending on the particular experimental conditions. Numerous experimental observations in the literature are in agreement with this interpretation. However, if the vortex lattice melting is indeed always responsible for the first order transition, all experimental results  must be consistent with this point of view. Apparently, this is not the case. We see no way in which the negative values of $\Delta M$, clearly demonstrated in Figs. 9(a) and 9(b), may be explained by invoking the vortex-lattice-melting hypothesis. At present, it is difficult identify the reason for the negative $\Delta M$ values. Several possibilities are discussed in Ref. \citen{cond}.

\subsection{Analysis of $M(T)$ curves}

In this subsection, we consider an experimental study of a Bi-2212 sample presented in Ref. \citen{kimura}. This study was chosen because both $M(H)$ and $M(T)$ curves were measured in rather extended ranges of magnetic fields and temperatures. Our main goal here is to demonstrate that the scaling procedure can also be used for the analysis of magnetization curves measured in fixed magnetic fields. 

\begin{figure}[h]
\centerline{\psfig{file=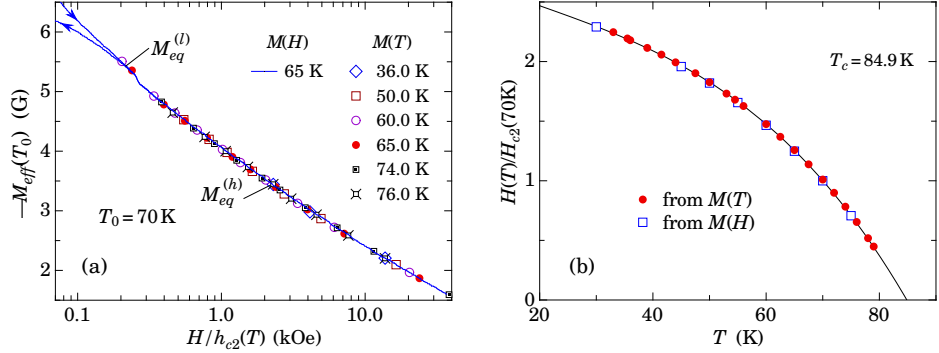,height= 1.9 in}}
\caption{(a) Comparison of $M_{eff}(H/h_{c2})$ for a Bi-2212 sample investigated in Ref.   \protect\citen{kimura} as calculated from $M(T)$ data (symbols) and as calculated from $M(H)$ data (solid line). (b) The resulting $h_{c2}(T)$ curve. The solid line represents the "universal" $h_{c2}(T)$ curve shown in Fig. 6(a), calculated for $T_c = 84.9$ K and $T_0 = 70$ K.}
\end{figure}
Because only $M(H)$ curves can be used for the evaluation of the scaling parameters $h_{c2}(T)$ and $c_0(T)$, the first step of the analysis is to transform the $M(T)$ curves into $M(H)$ data at different temperatures. It can easily be done by taking the corresponding values directly from the $M(T)$ curves. The correspondingly scaled magnetization is presented in Fig. 10(a). Different  data sets perfectly merge together, allowing for a reliable evaluation of $h_{c2}(T)$ and $c_0(T)$. As may be seen in Fig. 10(b), the values of $h_{c2}(T)$  that emerge from the scaling of either the $M(T)$ or the $M(H)$ curves coincide. 

\begin{figure}[h]
\centerline{\psfig{file=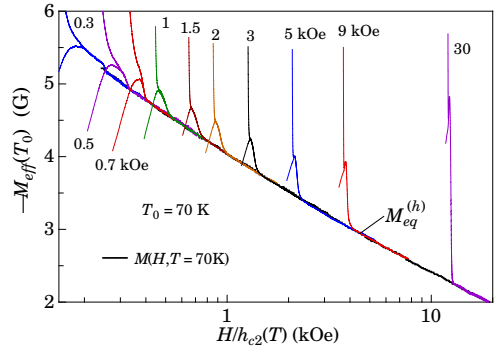,height= 1.85 in}}
\caption{(a) The $M(T)$ data, as reported in Ref. \protect\citen{kimura} for Bi-2212, scaled as described in the text. The $M(H)$ curve measured at $T = 70$ K is shown for comparison as a thicker solid line. }
\end{figure}
Although the algorithm described above allows for establishing the temperature dependence of the normalized upper critical field $h_{c2}(T)$, the small number of data points along the resulting $M(H)$ curves makes them ineffective for the analyses of the phase transition itself. In order to make use of all the data, the following procedure may be applied. After  $h_{c2}(T)$ and $c_0(T)$ are established, the transformation defined by Eq. (3) can be applied to each point of each of the original $M(T)$ curves. The results are plotted in Fig. 11 as $M_{eff}(T_0)$ versus $H/h_{c2}(T)$. Note that the temperature variation of $M$ enters via $h_{c2}$. For temperatures $T < 35$ K, extrapolations of the  $h_{c2}(T)$ and $c_0(T)$ curves were used. The measured isothermal $M(H,70$K) curve is shown for comparison. As may be seen, above the transition the $M_{eff}(H,T_0)$ curve, calculated from the $M(T)$ data collected in different applied fields, coincides with experimentally collected isothermal magnetization at $T = T_0$, i.e., it does not really matter whether the change of $H/h_{c2}(T)$ is achieved by varying the applied magnetic field or by a corresponding change of temperature. 

Because the magnetization curves above the transition are reversible, the equilibrium magnetization curve for the high-field phase $M_{eq}^{(h)}$ may reliably be established in a rather extended range of $H/h_{c2}$. In order to compare the transitions observed at different magnetic fields, we plot the difference $M_{eff} - M_{eq}^{(h)}$ in Fig. 12(a). In view of these plots, we argue that substantial changes of $M$ in rather narrow ranges of $H/h_{c2}(T)$ provide convincing evidence for the thermally induced phase transitions in all available magnetic fields. The temperature dependence of the mid-point of the transition $H^*$ is shown in Fig. 12(b). It turns out that $H^*(T)$ may very well be approximated by a power law for $T \ge 45$ K. At lower temperatures, $H^*$ increases exponentially with decreasing temperature. Both these features are emphasized by the broken and the solid line, respectively.
\begin{figure}[h]
\centerline{\psfig{file=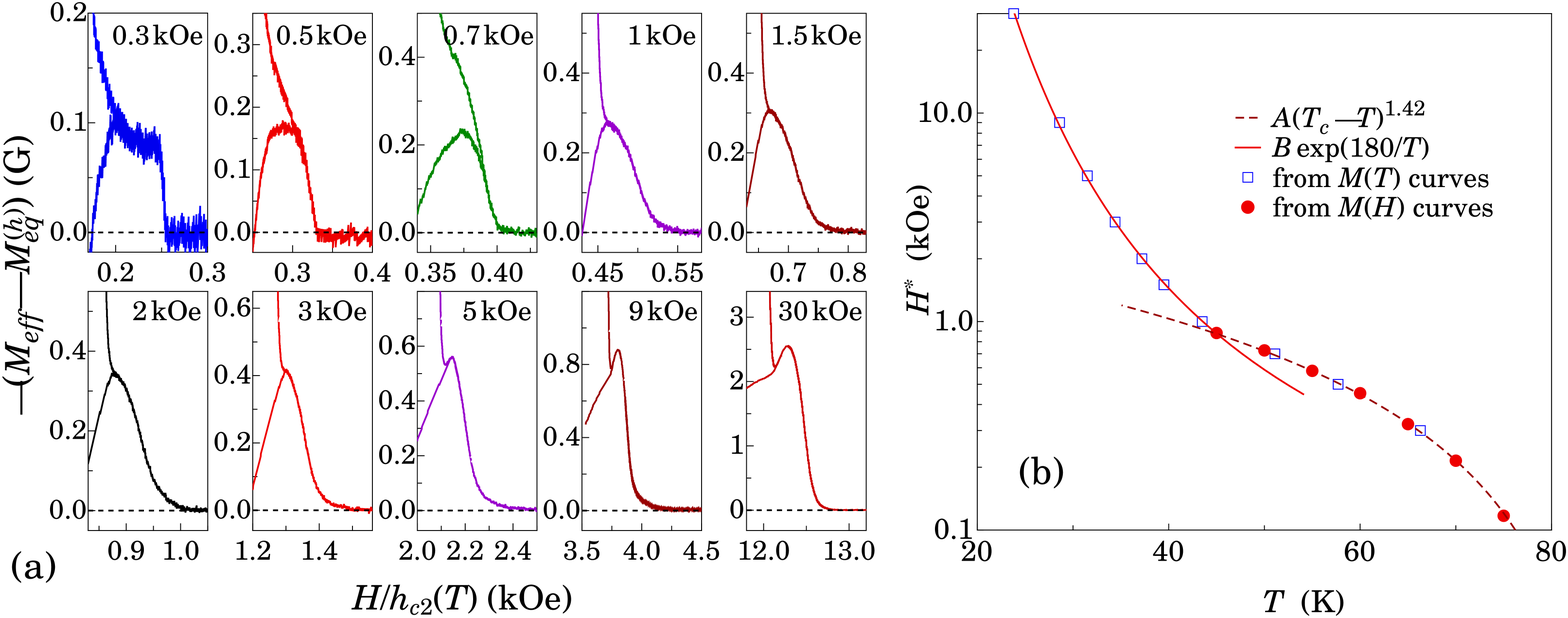,height= 2. in}}
\caption{(a) The difference $M_{eff} - M_{eq}^{(h)}$ as a function of $H/h_{c2}$. (b) The middle-point of the transition as a function of temperature. The solid and the broken line are commented in the text; $A$ and $B$ are fit parameters.}
\end{figure}

The experimental results that we discuss here are available in the literature for already some time and several scenarios for their interpretation have been proposed.\cite{kimura,feng,kobl,wen} There is a consensus that in the high temperature (low field) range, the step-like feature in the magnetization curves reflects the vortex-lattice-melting transition. The situation at low temperatures and high fields is still controversial. Nevertheless, it seems to be commonly accepted that the phase transition at low temperatures is {\it not} reflecting the melting of the vortex lattice. While our analysis of experimental data presented above cannot serve to identify the exact nature of the phase transition, it may be used to verify the validity of different models. Below we consider different explanations of the magnetization step in the low-temperature regime that we could find in the literature and show that all these explanations are inconsistent with the existing experimental data.  

1. In Ref. \citen{kimura} it was proposed that with decreasing temperature the first order vortex-lattice-melting transition gradually changes into a second-order vortex glass transition. In a more recent study,\cite{feng} a similar behavior in the low temperature range was attributed to a second order transition between two different vortex liquids. Neither of two can be accepted. As may be seen in Fig. 12(a), there are equally sharp and large changes of the reversible magnetizations in all applied magnetic fields, clearly a feature of a first rather than a second order phase transition.  

2. In Ref. \citen{kobl}, it was assumed that the $H^*(T)$ curve at lower temperatures represents the $H_{c2}(T)$ curve for inclusions with a lower critical temperature in a chemically non-uniform sample. However, considering the exponential increase of $H^*$ with decreasing temperature at $T < 45$ K, this explanation must inevitably be rejected. 

3. Another type of spatial non-uniformity of the investigated sample was considered in Ref. \citen{wen}. It was assumed that this non-uniformity is induced by the so-called intrinsic inhomogeneous electronic state in HTSC's. In this scenario, two different phases of the mixed state above $H^*(T)$ and an additional phase boundary separating them are expected. However, our scaling analysis presented above reveals only one single phase of the mixed state above $H^*(T)$ and therefore this explanation is also not tenable. 

Summarizing this part of the discussion, we note that to the best of our knowledge, no acceptable explanation of the magnetization step in the mixed state of HTSC's in high magnetic fields has yet been given. This situation is all the more disturbing, because this step is a rather prominent feature of the magnetization curves.


\begin{thebibliography}{65}
\bibitem{fn0} We note that these arguments are not completely justified. A short discussion of this matter may be found in Ref. \citen{1}
\end{thebibliography}
\end{document}